\documentclass[aps,twocolumn,prb,amsfonts,showpacs,showkeys,footinbib,superscriptaddress,byrevtex,a4paper,citeautoscript]{revtex4-1}
\usepackage[dvips]{graphicx}
\usepackage{dcolumn}
\usepackage{lineno}

\begin{document}

\title{Electronic and atomic structure of complex defects in highly n-type doped ZnO films}
\author{E. \surname{Men\'{e}ndez-Proupin}}
\affiliation{Instituto de Energ\'ia Solar and Dept. Tecnolog\'ias Especiales, E.T.S.I. Telecomunicaci\'on, Universidad Polit\'ecnica de Madrid, Spain}
\affiliation{Departamento de F\'isica, Facultad de Ciencias, 
Universidad de Chile, Las Palmeras 3425, 780-0003 \~Nu\~noa, Santiago, Chile}
\author{P. Palacios}
\affiliation{Instituto de Energ\'ia Solar and Dept. Tecnolog\'ias Especiales, E.T.S.I. Telecomunicaci\'on, Universidad Polit\'ecnica de Madrid, Spain}
\affiliation{Instituto de Energ\'ia Solar \& FyQATA, E.I. Aeron\'autica y del Espacio, Universidad Polit\'ecnica de Madrid}
\author{P. \surname{Wahn\'on}}
\affiliation{Instituto de Energ\'ia Solar and Dept. Tecnolog\'ias Especiales, E.T.S.I. Telecomunicaci\'on, Universidad Polit\'ecnica de Madrid, Spain}
\date{\today}

\begin{abstract}
Point defects in Ga- and Al-doped ZnO thin films are studied by means of first principles electronic structure calculations. Candidate defects are identified to explain recently observed differences in electrical and spectroscopical behavior
of both systems. Substitutional doping in Ga-ZnO explain the metallic behavior of the electrical properties. Complexes of interstitial oxygen with substitutional Ga can behave as acceptor and cause partial compensation, as well as gap states below the conduction band minimum as observed in photoemission experiments. Zn vacancies can also act as compensating acceptors.  On the other hand, the semiconducting behavior of Al-ZnO and the small variation in the optical gap compared with pure ZnO, can be explained by almost complete compensation between acceptor Zn vacancies and substitutional Al donors.  Interstitial Al can also be donor levels and can be the origin of the small band observed in photoemission experiments below the Fermi level. Combinations of substitutional Al with interstitial oxygen can act simultaneously as compensating acceptor and generator o the mentioned photoemission band.  The theoretical calculations have been done using density functional theory (DFT)  within the generalized gradient approximation with on-site Coulomb interaction.  In selected cases, DFT calculations with semilocal-exact exchange hybrid functionals have been performed. Results explain  photoelectron spectra  of Ga-ZnO and Al-ZnO at the corresponding doping levels.
\end{abstract}
\pacs{}
\keywords{}
\maketitle

\section{Introduction}

Doped ZnO based materials  constitute a family of transparent conducting oxides with several potential applications in solar cells, 
windows thermal coatings and spintronic devices\cite{hanada}. N-type doping, easily achieved with group III elements (Al, Ga, In,...)-although 
In does not seem to be an economical and nature-friendly option due to its natural scarcity and toxicity-, improves both their electrical and 
optical properties.\cite{minami05} However, and in spite of the considerable effort that scientific community is doing in order to 
understand the mechanisms that rule the doping effectiveness in such ZnO doped materials, there is still a lack of knowledge 
on the dopants interaction with ZnO host matrix and their intrinsic defects and how it affects to the ZnO electronic structure. 

 Recent results on Ga- and Al-doped ZnO thin films 
grown by magnetron sputtering (doping content 1\% at. for both films) showed substantial differences in their electric and optical properties.\cite{gabas1} 
On the one hand, 
Ga-ZnO temperature resistivity behavior was metallic-like and the film presented an increased optical bandgap, 
3.63 eV vs. 3.21 eV measured for the undoped film, both facts being consistent with substitutional doping.
On the other hand, Al-ZnO film behaved as a semiconductor and showed little variation in the optical gap (3.25 eV) compared with pure ZnO. 
 Since Al and Ga have similar electronic structure in their valence levels, they were expected to 
behave analogously as substitutional dopants in the ZnO matrix. Hence, the different doped film behaviors were then attributed to 
the tendency of Al and Ga cations to occupy different insertion sites in the host ZnO.

The electrical properties of ZnO and Al-ZnO correspond to a semiconductor with conduction electrons thermally activated, i.e.,  the resistivity decreases with increasing temperature. 
The resistivity of Al-ZnO ($3.6-2.6\times 10^{-2}$  $\Omega$~cm)  
is one order of magnitude smaller than for undoped ZnO.  
 Ga-ZnO presents metallic behavior, with resistivity increasing with temperature
  ($9.7-9.85 \times 10^{-4}$ $\Omega$~cm), 
 about 30 times smaller than for Al-ZnO. The Ga-ZnO film carrier density 
 deduced from Hall effect measurements\cite{gabas1} is one order of magnitude larger than for Al-ZnO
 ($5\times 10^{20}$ vs $3\times 10^{19}$ cm$^{-3}$). 
 From these carrier concentrations, a conduction band population of 0.7 and 0.04 electrons per dopant atom  can be inferred in Ga-ZnO and Al-ZnO, respectively.  Therefore, doping effectiveness seems to be modulated by the presence of acceptor defects,
  that would compensate partially (Ga-ZnO) or almost totally (Al-ZnO), the donor doping.  

New data\cite{gabas2}   from hard X-ray photoemission spectroscopy (HAXPES) 
have revealed an electronic band in the doped material, near the conduction band 
minimum (CBM), which is considerably stronger in Ga-ZnO than in Al-ZnO. HAXPES 
is better suited than conventional photoelectron spectroscopy for the exploration of 
the  density of states (DOS) at the valence band (VB) and Fermi level  regions, since 
the contribution of surface features is strongly reduced.\cite{Sacchi2005, Panaccione2012} 
 
 In this Article, we explore the electronic structure of a number of defects in heavily doped Al-ZnO and Ga-ZnO by means of density functional theory (DFT) calculations. We identify the defects that can explain the peculiar HAXPES band and the electrical properties observed in Al-ZnO and Ga-ZnO. 
The defects of ZnO have been studied intensively using DFT in recent years.\cite{kohan2000,zhang01,erhart05,janottivandewalle05,lanyzunger07prl,janotti07,hse0375,oba2011}  
These studies has focused on the thermodynamical properties and the electronic structure of isolated defects, and none can explain the observed HAXPES band. 
Our study is focused at Al-ZnO and Ga-ZnO with $\sim 1$~\% at. concentration of Al or Ga. 
The Article is organized as follows. The computational methods are explained in Sect. \ref{sec:methods}, the computed defects electronic structures are presented  in Sect. \ref{sec:results}. Sect. \ref{sec:conclusions} is devoted to our conclusions.

\section{Methods}
\label{sec:methods}

The local density approximation (LDA) and the generalized gradient approximation (GGA) are the most commonly used flavors of DFT. Their greatest limitation for semiconductor materials is the underestimation of the fundamental bandgap.  In the case of ZnO, part of this inaccuracy is produced because  the Zn 3d 
binding energy is underestimated by several eV due to the self-interaction error. Therefore, the Zn 3d and O 2p levels, present an incorrectly large hybridization, pushing up the top valence band composed mainly of O 2p levels.\cite{janottivandewalle06}  The on-site Coulomb interaction method (GGA+U)\cite{dudarev} allows to obtain the correct binding energy for the Zn 3d levels, using 
a Hubbard term correction for the 3d levels of Zn with the parameter $U-J=8.5$ eV\cite{palaciosznoal09,palaciosznoal2010}). Correcting the energy of Zn 3d states, 
 the mixing with O 2p states is reduced, and the bandgap values are improved, although not totally. Most of our calculations have been made using the GGA+U method. A plane-wave projector augmented wave\cite{paw1,paw2} scheme has been used, as implemented in the Vienna Ab Initio Simulation Package (VASP)\cite{vasp4}. The GGA exchange-correlation functional of Perdew, Burke, and Ernzerhof (PBE)\cite{pbe} has been used.  

For selected cases, we have used the hybrid functional of Heyd, Scuseria and Ernzerhof (HSE)\cite{hse,hse06}. This functional generally allows to obtain better bandgaps and better structural properties than PBE, at the cost of a great increment in computer time. Following the recent practice\cite{hse0375,oba2011,demchenko2011} we have used the fraction 0.375 of the Hartree-Fock exact exchange, that allows to fit the experimental gap of ZnO, 3.4 eV. The DOS of substitutional Ga shown in Fig. \ref{fig:xps-gap} was obtained within this approximation. 

The primitive unit cell of ZnO have four atoms. To simulate the impurity concentration 1~\% at., we have used a $3\times 3\times 3$ supercell, containing 108 atoms (109 and 107 atoms in case of interstitials and vacancies, respectively).
A plane wave cutoff of 500 eV, was used in all GGA+U calculations and HSE calculations with the ZnO unit cell. The lattice constants used were $a=3.2473$~\AA, $c=5.2085$~\AA, as obtained from structural relaxation with the HSE functional. For HSE  calculations with the $3\times 3\times 3$ supercell, a reduced cutoff of 400 eV was used in order to decrease the computational time. 
 With 400 eV, the pressure is underestimated by 39 kbar, but it may be safely used for simulations with constant cell.
The Brillouin zone was sampled with  $3\times 3\times 2$, and $6\times 6\times 4$ $\Gamma-$centered k-point grids for structural optimization and for DOS calculations, respectively. 
Defect images were made with the Visual Molecular Dynamics (VMD) software.\cite{vmd}

\section{Results and discussion}
\label{sec:results}

\begin{figure}[!]
\includegraphics[width=8.5cm]{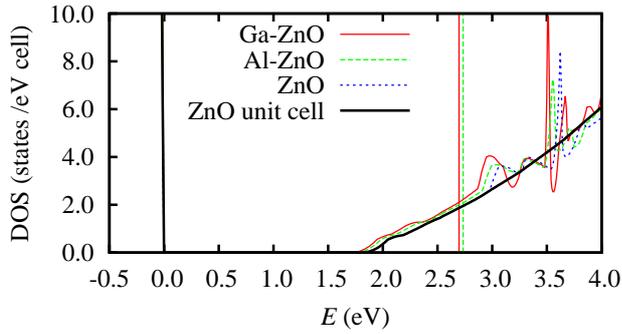}
\caption{(Color online) Bottom: DOS around the fundamental bandgap. The red and green vertical lines indicate the Fermi levels of Al-ZnO and Ga-ZnO.
The almost vertical black line at 0 eV is the VB edge.
\label{fig:dos-cationsust}}
\end{figure}

Fig. \ref{fig:dos-cationsust} shows the DOS near the fundamental bandgap for undoped ZnO and  Al- and Ga-doped ZnO, 
assuming that all dopant cations are in substitutional sites. 
All DOS have been computed using the same $3\times 3\times 3$ supercell (108 atoms), PBE exchange-correlation functional, 
and other  
computational parameters. Also shown is the ZnO DOS computed with the 4-atoms unit cell and a denser $18\times 18\times 12$ k-points grid 
that should be equivalent to the coarse grid used with the supercell. The oscillations 
observed in the supercell DOS above the Fermi 
level are an artifact of the interpolation of energies in the Brillouin zone (tetrahedron method), and they should be reduced using a 
denser k-points grid. However, these oscillations  have no effect in the occupied states. It is seen that the doping-induced change 
on the conduction band (CB) DOS is minimal, both dopant cations supply 
an extra electron that populates a perturbed host state at the bottom of the conduction band, but there is no differentiated band below Fermi level associated with the Al and Ga
 cations occupying substitutional positions in ZnO matrix. 
Due to the Pauli exclusion principle, the optical bandgap must 
renormalize by the difference between the Fermi level and the valence band maximum (VBM).\cite{burstein54,moss54} 
 
The theoretical gap values obtained are $E_g$(ZnO)=1.806 eV, $E_g$(Ga-ZnO)=2.696 eV, $E_{cv}$(Ga-ZnO)=1.709 eV, 
$E_g$(Al-ZnO)=2.734 eV, $E_{cv}$(Al-ZnO)=1.745 eV. $E_{cv}$ is the gap between the VBM and the CBM. 
The band-filling energy is $\Delta E_{bf}=0.989$ and 0.987 eV (equal within error) for Al and Ga doping. 
If each impurity donated one electron to the CB, 
the optical gap variation 
would be 0.93 and 0.89 eV for Al- and Ga-ZnO, respectively. However, it is found experimentally that Ga-ZnO increases its 
gap by only 0.42 eV, while for Al-ZnO the increment is 0.04 eV.\cite{gabas1}
 Therefore, the dopant cations have to be partially 
compensated by acceptor defects that would reduce the carrier density at the conduction band. Thereafter, we need to find 
models of shallow acceptors associated to Al and Ga dopants. 

The above mentioned Fermi levels, gaps and band-filling energies have been obtained from supercell calculations with 
108 atoms, i.e, 0.926 \% at. impurity concentration. For our supercells containing just one impurity atom the  
conduction band is populated by exactly one electron. As the dopant experimental concentration is nominally 1 \% at., and it is not 
feasible to perform calculations with 100-atoms supercells we need to correct the band-filling energies. We find the 
corrected Fermi level ($E_F^{\prime}$) requiring that the conduction band population is 1.08 electron. The renormalized 
band-filling energies are given as $\Delta E_{bf}^{\prime}=E_F^{\prime}-E_{cv}$. We are neglecting changes in $E_{cv}$
assuming that it depends more on the nature of the impurity than on its concentration. This is not completely 
justified because the impurity states are shallow and extend over the full supercell.  Considering that replacing one Zn atom by Ga causes 
a change of $-0.1$ eV in $E_{cv}$, and assuming a linear dependence of $E_{cv}$ with the impurity concentration, the correction would be
 -7.4 meV. Hence, we estimate a correction for $\Delta E_{bf}^{\prime}-\Delta E_{bf}$  of $0.03$ eV. 

Let us consider carrier concentrations inferred from transport data, i.e., 0.7 and 0.04 electrons per Ga and Al atom, respectively. Renormalizing these values by the factor 1.08, we find effective band-filling energies of 0.86 eV (Ga-ZnO). For  Al-ZnO, the band-filling effect cannot be estimated with our rough k-points sampling. Using the effective conduction and heavy valence masses of ZnO (0.22 $m_e$ and 3.03 $m_e$\cite{hanada}), the band-filling energy is 0.14 eV. Adding $E_{cv}$ we obtain an effective gap of 1.88 eV, which is 0.07 eV larger than for intrinsic ZnO. This difference is in good agreement with the measured optical gaps.

There are two possible explanations for the reduced gap shift in Ga-ZnO. One may be the band-gap narrowing effect, a many body effect that is not accounted completely by DFT calculations\cite{persson2009}. This effect is produced for carrier densities larger than 
a critical value, and is associated with the insulator-metal Mott transition. 
Other possibility is that the band 
observed in HAXPES is due to a non substitutional defects. This band could be partially filled and be responsible of conductivity. 

\begin{figure}[!]
\includegraphics[width=8.5cm]{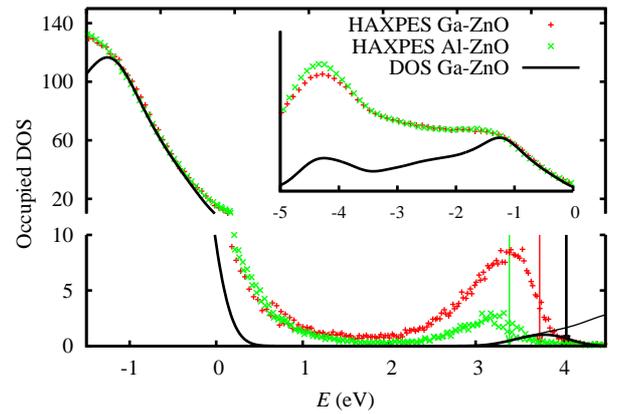}
\caption{(Color online) Theoretical DOS around the fundamental gap, computed for the $\mathrm{Ga_{Zn}}$ defect using the gap-corrected HSE(0.375) method. Also shown are the HAXPES spectra for Al-ZnO and Ga-ZnO, data by courtesy of M. Gab\'as, P. Torelli, N. Barrett and M. Sacchi.
\label{fig:xps-gap}}
\end{figure}

\begin{figure*}[!]
\includegraphics[width=16.0cm]{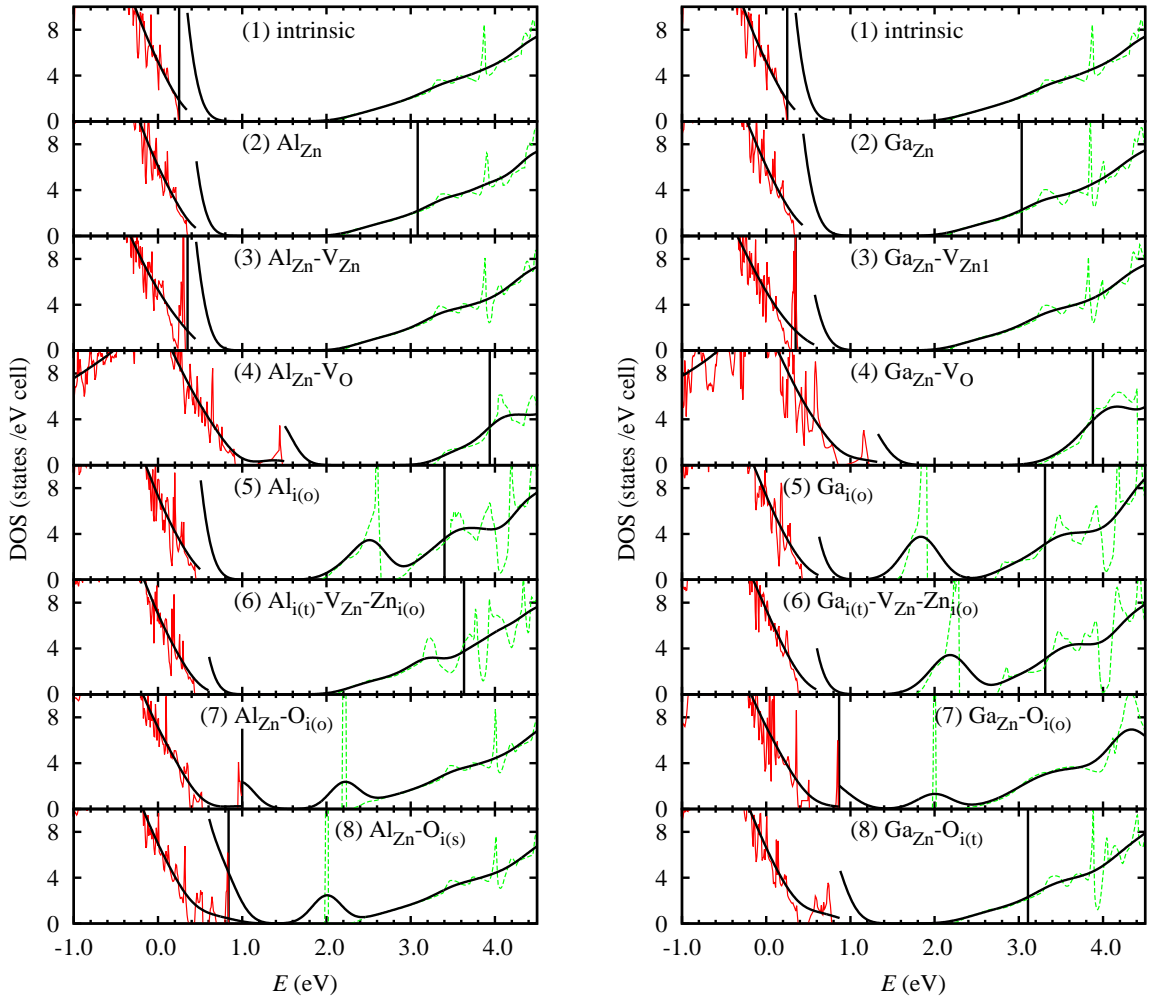}
\caption{(Color online) DOS of different doping situations considered for Al and Ga doping. The valence band DOS is shown divided by 10 for optimal view. Also shown, in thick black line, the DOS smeared with the experimental HAXPES resolution (see text).
\label{fig:dos-modelos}}
\end{figure*}

Figure~\ref{fig:xps-gap} shows the HAXPES spectra of Al-ZnO and Ga-ZnO in the region of the fundamental bandgap. The theoretical DOS calculated for Ga-ZnO is shown for comparison, assuming that all the dopant cations are placed in substitutional sites, $\mathrm{Ga_{Zn}}$, 
replacing thus the Zn cations. This DOS has been computed with the HSE(0.375) functional, in order to be free of the gap error and allow a quantitative comparison with the HAXPES results.
The energy scale zero has been set to the theoretical VBM, and the HAXPES spectra have been aligned to the DOS by the VB edge.  The inset in Fig.~\ref{fig:xps-gap} shows that not only the VB edge, but the full bands coincide when they are aligned in this way.
The HAXPES spectral function is roughly proportional to the DOS multiplied by the 
Fermi-Dirac occupation function, and convoluted with a Gaussian-like distribution that 
accounts for the resolution of the photoelectron detector of the binding energies. 
Henceforth, to facilitate the comparison, the DOS 
has been multiplied by the Fermi-Dirac occupation function, and broadened with a Gaussian function with standard deviation $\sigma=0.206$~eV.\cite{haxpesfit} 
 Some important effects neglected in this approximation are the quantum transitions and photoelectron escape probabilities, which affect the energy-dependent ratios of the DOS to HAXPES signal, as illustrated in the inset for the full range of the VB.  The position of the Fermi level modifies the HAXPES spectrum at the Fermi level. In the experiments, the Fermi level is fixed by the gold contacts in electrical equilibrium with the semiconductor. The alignments of the semiconductor bands  with respect to the common Fermi level depend on the carrier concentration, and the formation of a Schottky barrier at the metal/semiconductor interface. 
The Fermi levels are indicated by vertical lines in Fig.~\ref{fig:xps-gap}. The Fermi level for the theoretical Ga-ZnO DOS (4.05 eV) corresponds to a CB population of 0.7 electrons per every 100 atoms. The corresponding band-filling is 0.88~eV, quite similar to the GGA+U value 0.86 eV above mentioned. 
The DOS without the occupation function is also shown in black thin line, as can be seen for energies larger than 3.7 eV. 
 
 The Ga-ZnO total gap obtained with HSE(0.375) is 4.05 eV, which is 
 0.65~eV higher than for pure ZnO. Comparing with the experimental  gap  shift,  0.42 eV, a many-body gap narrowing of 0.23 eV can be inferred. Following this reasoning, the optical gap would equal the difference between the Fermi level and the VBM, 3.63 eV. For Al-ZnO, where the many-body gap narrowing is  absent because it is not a metal, the same difference is 3.25 eV.  The difference between the gaps of Ga-ZnO and Al-ZnO is 0.38 eV, in close agreement with the difference between the experimental Fermi levels, 0.35 eV.\cite{alignment} To apply the gap-narrowing correction in Fig.~\ref{fig:xps-gap}, it is enough to redshift  the DOS curve for energy above 3 eV.
 
From Fig.~\ref{fig:xps-gap}, it is evident that Ga-ZnO presents more states than Al-ZnO between 2.5 and 3.5 eV, and the ratio of areas is not proportional to the ratio of free carrier densities, which is one order of magnitude smaller for the Al doped films.    
If these peaks were due to the occupation of the conduction band by the electrons supplied by the impurities, one would expect some coincidence at the
 low energy side of the peak between 2.5 and 3.5 eV. Moreover, the  width of this peak for Al-ZnO is inconsistent with the 
 experimental low amount of free CB  electrons. Hence, there must be some defects that cause 
localized (non-conducting) states just below the CB edge. Comparing the  HAXPES signals with the DOS for the ranges  $0-1$ eV and $2.5-3$ eV, the presence of wide band tails can be inferred in both Al-ZnO and Ga-ZnO, which cannot be attributed to the HAXPES resolution, but to the presence of real electronic states. 

In order to explain the HAXPES signal, we will investigate  the types of defects  that can cause it, as well as the acceptor defects that take electrons out of the CB.

Figure \ref{fig:dos-modelos} shows the DOS calculated for a number of defect combinations in the ZnO matrix, mostly assuming that the 
dopant cations substitute Zn cations and assuming the neutral charge state. Charged states will be discussed separately. The possibility of the dopant cations as interstitial impurities is also shown. 
To facilitate the visualization, the DOS of the valence band has been divided by 10 (red line).
The highest occupied level for each model are indicated by  black vertical lines.  Thick black lines show the broadened DOS according to experimental resolution. 
The jumps seen in the broadened DOS are due to the above mentioned the factor $1/10$ for the valence DOS. The Fermi-Dirac occupation function has not been used for this figure, because the Fermi level is unknown. 
As the Fermi level cannot be determined accurately, we prefer to analyze the full DOS. However, considering the previous discussion, one must keep in mind that the Fermi level is near the CBM for Al-ZnO, and approximately 0.86 eV over the CBM for Ga-ZnO. 
Fig. \ref{fig:dos-modelos} (1) show the DOS of intrinsic ZnO in order to facilitate the comparison with the defect models. Fig. \ref{fig:dos-modelos} (2) show the DOS for substitutional cations, that has already being discussed. 

\subsection{Substitutional doping with Zn vacancies}

\begin{figure}[!]
\includegraphics[width=8.0cm]{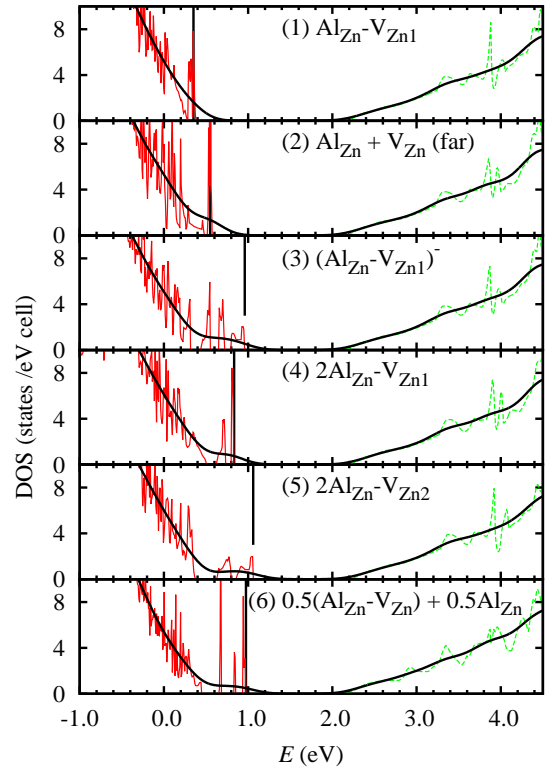}
\caption{(Color online) DOS of different configurations of the complex of Zn vacancies with substitutional Al.
\label{fig:alzn-vzn}}
\end{figure}

The substitutional-vacancy complex $\mathrm{M_{Zn}-V_{Zn}}$ (M=Al, Ga) presents a shallow acceptor band about 0.1 eV over the valence 
band  (Fig. \ref{fig:dos-modelos} (3)). This level can accept one electron and compensate the donor $\mathrm{M_{Zn}}$. The configuration with the Zn 
vacancy in the same (001) plane as the substitutional  Al is 0.08 eV more stable than the case where the vacancy and the Al cation are placed in adjacent 
planes. The presence of Zn vacancies is favored from the thermodynamical point of view, as the Zn vacancy has the lowest formation energy 
in O-rich conditions, when the Fermi level approaches to the CBM\cite{lanyzunger07prl,demchenko2011}. The vacancy can be in the same atomic plane as the cation, or in an adjacent plane, which is 0.08 eV higher in energy, but has almost the same DOS. Fig. \ref{fig:alzn-vzn} shows the DOS for different 
configurations of this defect, including charged states. 

For the sake of clarity, before continuing the description of other dopant defect combinations in Fig.~\ref{fig:dos-modelos}, we present in 
Fig. \ref{fig:alzn-vzn} the DOS of different configurations of the vacancy-cation complex. 
Fig. \ref{fig:alzn-vzn}(1) shows the DOS for Al and vacancy located at adjacent 
atomic planes. This 
configuration is indicated by the subindex 1. Subindex 2, indicates that the cation and the vacancy are in the same plane. 

The vacancy could be separated from $\mathrm{M_{Zn}}$, but in this case 
the formation energy is somewhat higher and  the acceptor band is shifted away from the VBM, as seen in Fig. \ref{fig:alzn-vzn}(2) for M=Al. 
The results for Ga-ZnO are practically the same and are not shown.   
When the Fermi level is near the CBM, the complex  $\mathrm{M_{Zn}-V_{Zn}}$ accepts one electron and is denoted  $(\mathrm{M_{Zn}-V_{Zn}})^{-}$. 
Its DOS is shown in Fig. \ref{fig:alzn-vzn}(3). 
The triple 
complex of two substitutional cations and a Zn vacancy $\mathrm{2M_{Zn}-V_{Zn}}$  has  the donor level separated from the 
VBM, but is within the range of the tail observed in the HAXPES spectra. $\mathrm{2M_{Zn}-V_{Zn1}}$  (Fig. \ref{fig:alzn-vzn}(4))  
corresponds to both $M$ atoms in the same 
basal plane and the vacancy in the adjacent plane, while $\mathrm{2M_{Zn}-V_{Zn2}}$  (Fig. \ref{fig:alzn-vzn}(5)) has both M atoms in adjacent 
planes and the vacancy in one of them. 
This defects with two cations have been calculated using the same 
supercell as the defects with one cation. Hence, it strictly corresponds to 1.8 \% at. concentration. 
For a double supercell, we expect the DOS to 
be an average between the DOS of intrinsic ZnO and $\mathrm{2M_{Zn}-V_{Zn}}$, or between $(\mathrm{M_{Zn}-V_{Zn}})^{-}$ and $\mathrm{M_{Zn}}^{+}$. 
 This behavior is in fact observed comparing the defect named 
$\mathrm{0.5(M_{Zn}-V_{Zn})+0.5M_{Zn}}$  (Fig. \ref{fig:alzn-vzn}(6))   with the charged $\mathrm{(M_{Zn}-V_{Zn})^{-}}$  (Fig. \ref{fig:alzn-vzn}(3))   and the uncharged 
$\mathrm{M_{Zn}-V_{Zn}}$  (Fig. \ref{fig:alzn-vzn}(1)). The defect (6) has been calculated using a double supercell than contains 
$\mathrm{M_{Zn}-V_{Zn}}$ far from $\mathrm{M_{Zn}}$, and the index 0.5 summed indicate the same concentration of substitutional dopants as the single 
supercells. In this case, it is seen that the valence band edge  has the same forms of those of 
the charged $\mathrm{(M_{Zn}-V_{Zn})^{-}}$, the charge coming from the donor $\mathrm{M_{Zn}}$. 
In practice, this model is nearly indistinguishable from the defect  $\mathrm{2M_{Zn}-V_{Zn}}$.
Hence, more important than the precise composition is to have the acceptors and donors in the correct charge state. 
Other configurations are possible, e.g.,  both M atoms and the vacancy in three different basal planes, but are not considered here due to the similarities
of the DOS for all the models.  

Based on the above discussion, one may design a model of compensating defects $\mathrm{M_{Zn}-V_{Zn}}$ and $\mathrm{M_{Zn}}$, that can be either close or far, where the first defect accepts one electron donated by a far
 $\mathrm{M_{Zn}}$. The vacancies can be also isolated, as in Fig.~\ref{fig:alzn-vzn}(2), but their total energy is higher and the vacancy tends to migrate towards the substitutional atom.

 In this model, the difference between Ga- and Al-doping would be the concentration of Zn vacancies, resulting in a 
 different degree of compensation. About half of Al is in the form $\mathrm{Al_{Ga}-V_{Zn}}$ hence attaining an 
 almost total compensation of the donor $\mathrm{Al_{Ga}}$. In the case of Ga-ZnO, the concentration 
 of $\mathrm{Ga_{Zn}-V_{Zn}}$
 should be much smaller than the concentration of $\mathrm{Ga_{Zn}}$, in such a way that 0.7 electrons per dopant atom remain 
 uncompensated and free to conduct. If 15 \% of Ga is in the acceptor form $\mathrm{Ga_{Zn}-V_{Zn}}$, 
 and 85~\% is in donor form $\mathrm{Ga_{Zn}}$, then the conduction band will be populated by 0.7 electron per Ga impurity. 
 
However, the above model alone cannot explain the observation of states below the CB by HAXPES. A possibility is to include in the description interstitials and oxygen vacancies.
 
\subsection{Substitutional doping with O vacancies}
Oxygen vacancy ($\mathrm{V_O}$) is an abundant point defect in intrinsic ZnO, specially when it is grown in O-poor conditions.
In O-rich conditions, $\mathrm{V_O}$ and its complexes are not expected to attain significant concentrations because of its  high formation 
energy when the Fermi level is near the CBM.
Our calculations show that the complexes $\mathrm{M_{Zn}-V_O}$ (M=Al, Ga)  (Fig. \ref{fig:dos-modelos} (4))  in the neutral state supply one electron 
to  the CB. The DOS shows an in-gap band 0.2-0.4 eV over the VBM, that may contribute to the VB tail, and it shows no peak near the CBM.  
The charged state $(\mathrm{M_{Zn}-V_O})^{+}$ has a similar DOS (not shown), the CB gets empty, and the in-gap band remains near the VBM.
 
  The DOS of far defects ($\mathrm{M_{Zn}+V_O}$)  (not shown) 
  presents similar features, only shifting slightly the in-gap band to 1 eV over the VBM. This in-gap band is characteristic of the isolated 
  vacancy $\mathrm{V_O}$.\cite{hse0375} Being at 1 eV over the VBM, this should be observed in the HAXPES
  experiment, contrary to the evidence. States  $(\mathrm{M_{Zn}-V_O})^{2+}$ or with larger charge are not compatible 
  with the Fermi level observed in the experiments. 

\subsection{Interstitial doping}
\begin{figure}[!]
\includegraphics[width=8.0cm]{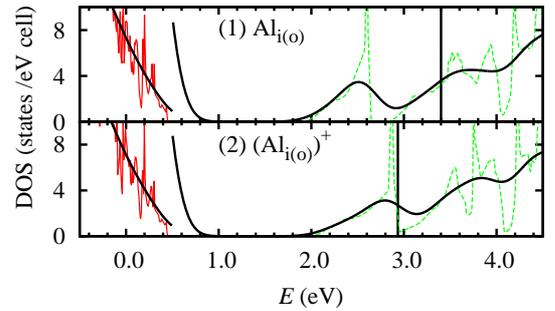}
\caption{(Color online) DOS of the neutral (1) and the positive singly charged (2) states  of the interstitial  $\mathrm{Al_{i(o)}}$. 
\label{fig:dos-ali}}
\end{figure}

Let us consider interstitial aluminum ($\mathrm{Al_i}$). 
At the octahedral interstitial site $\mathrm{Al_{i(o)}}$, the DOS shows a 0.7~eV wide band with a maximum 0.2 eV below the CBM
 (Fig. \ref{fig:dos-modelos} (5)). Such a defect may justify the small band observed in Al-ZnO HAXPES below the Fermi level 
 (see Fig.~\ref{fig:xps-gap}). It populates the CB with one electron. In order to match the experimental facts, 
 this defect needs to be positively charged  ($\mathrm{Al^{+}_{i(o)}}$), 
thus avoiding the filling of the CB and  the observation of two peaks in HAXPES. As can be seen in Fig. \ref{fig:dos-ali}, the charged state keeps an in-gap band totally filled and close to the CBM.

The corresponding Ga interstitial ($\mathrm{Ga_{i(o)}}$) presents similar features, but the in-gap band is deeper (0.8 eV below the CBM), 
and the CB is populated, as discussed above.  
This would produce a double peak structure in the HAXPES  Ga-ZnO spectrum that is not observed. Notice that the charged $(\mathrm{Ga_{i(o)}})^{+}$ is inconsistent with 
the observed population of the CB in Ga-ZnO.

 Tetrahedral interstitial ($\mathrm{Al_{i(t)}}$) is unstable and relaxes to a related configuration, that can be named $\mathrm{Al_{i(t)}-V_{Zn}-Zn_{i(o)}}$.  It turned out that the closest Zn was displaced from its lattice position 
 to a close octahedral site. Then Al moved  almost into the vacant lattice site, also attracting the closest O atom. In this configuration, 
 which has lower formation energy (0.6 eV less) than $\mathrm{Al_{i(o)}}$, the donor band overlaps with the conduction band, forming a 
 continuous DOS band 1.5~eV wide up to the Fermi level and populated with three electrons  (Fig. \ref{fig:dos-modelos} (6)) . 
 This is also in contradiction with the experimental electrical behavior. 
 
 The equivalent combination defect in Ga-ZnO has a somewhat different geometry due to the larger dopant 
 atomic radius. Compared with the $\mathrm{Al_{i(t)}-V_{Zn}-Zn_{i(o)}}$ defect, 
 the lowest conduction bands of $\mathrm{Ga_{i(t)}-V_{Zn}-Zn_{i(o)}}$ splits in a deep in-gap band 
 and a continuous upper band occupied by one electron. This should be seen in the HAXPES spectrum as a double band. 
 Therefore, this defect combination would  not explain the experimental facts. 

\subsection{Substitutional doping with O interstitials} 
Interstitial oxygen is thermodynamically favored
at O-rich conditions, only overcome by the Zn vacancies\cite{lanyzunger07prl,demchenko2011}. 
The different configurations have been analyzed in Refs.~\onlinecite{erhart05,janotti07}. The configuration with lowest formation energy 
for Fermi level between 0 and $\sim 2.8$~eV over the VBM, is the so called split\cite{janotti07} or dumbbell\cite{erhart05} 
configuration in neutral charge state $(\mathrm{O_{i(s)})^{0}}$. This configuration can be regarded as an O$_2$ dimer substituting a lattice O. 
This dimer is not spin-polarized (different to the free O$_2$) and the bond length is 1.49~\AA. 
For higher values of Fermi level, the charged octahedral configuration 
$(\mathrm{O_{i(o)})^{2-}}$ is more stable than the split configuration. 
Hence, the oxygen interstitial is a 
double acceptor, by means of a transformation from the split to the octahedral configuration.
According to Refs.~\onlinecite{erhart05,janotti07}, the singly charged state 
$(\mathrm{O_{i(o)})^{-}}$ 
is unstable in all configurations. Other metastable configuration is the split$\ast$ or rotated-dumbbell\cite{erhart05,janotti07}, which is 0.1-0.2 eV higher in 
energy. 

\begin{figure}[htbp]
\begin{center}
\includegraphics[width=7.0cm]{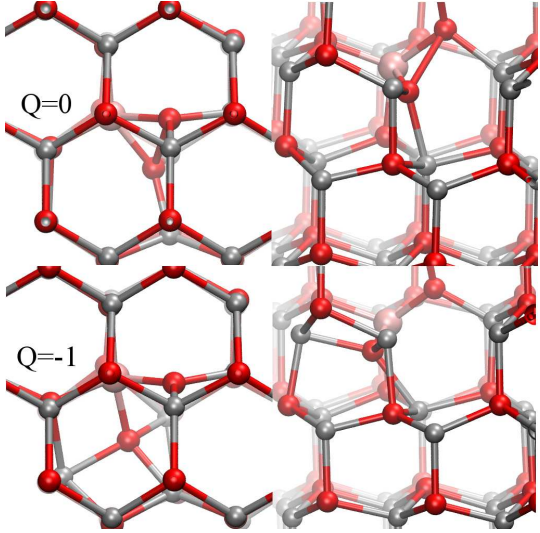}
\caption{(Color online) Views of defect $(\mathrm{Al_{Zn}-O_{i(o)}})$. Top: Neutral state. Bottom: state charged with one electron. Pink, red, and gray balls represent Al, O, and Zn atoms, respectively.}
\label{fig:defect-oi-alzn}
\end{center}
\end{figure}

\begin{figure}[!]
\includegraphics[width=8.0cm]{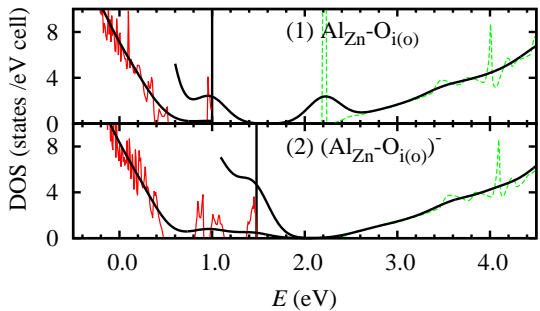}
\caption{(Color online) DOS of the neutral (1) and the negative singly charged (2) states  of the combination  $(\mathrm{Al_{Zn}-O_{i(o)}})$. 
\label{fig:dos-alzn-oi}}
\end{figure}

\begin{figure}[!]
\includegraphics[width=8.0cm]{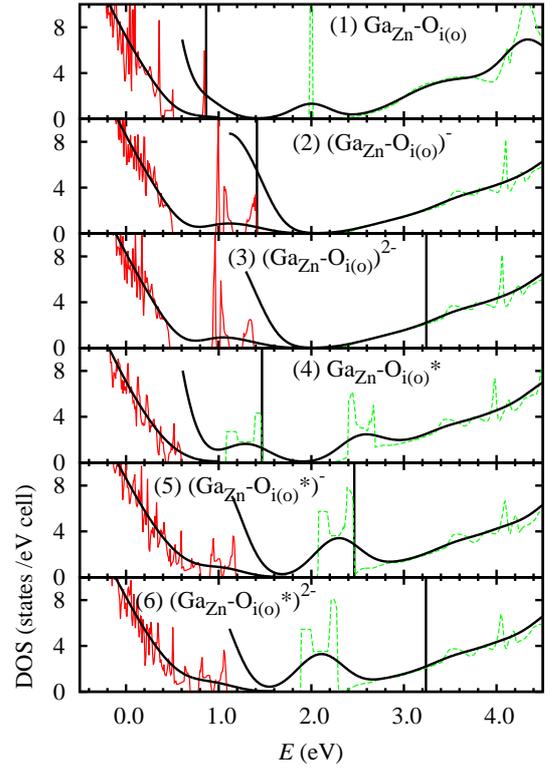}
\caption{(Color online) DOS of different configurations of the defect  combination  $(\mathrm{Ga_{Zn}-O_{i(o)}})$.
\label{fig:dos-gazn-oi}}
\end{figure}

In the proximity of a substitutional cation (M=Al, Ga), $\mathrm{O_i}$ may arrange in a split, split$\ast$, or octahedral configurations, or, considering the complex energy landscape of the isolated interstitial, take a different configuration. Its electrical behavior may be that of a single-acceptor, as result of the 
combination of a double acceptor with a single donor. 
 The lowest energy combination has been obtained relaxing the structure  from an octahedral interstitial position
    ($\mathrm{Al_{Zn}-O_{i(o)}}$) near the Al atom  (Fig. \ref{fig:defect-oi-alzn}). The neutral defect is more distorted than the charged one, 
    and it may be considered a kind of split of configuration, where the O$_2$ dimer has \textit{bond} length of 2.04 \AA{} and is oriented along the line joining 
    two octahedral cavities. In the charged state, the dimer breaks and the oxygens enter more in the cavities.  
Their DOS are shown in Figs. \ref{fig:dos-modelos}(7) and \ref{fig:dos-alzn-oi}.
   The neutral state is spin-polarized and its HOMO and LUMO have opposite spin 
  projection. The HOMO is 0.6 eV over the VBM, while the LUMO is just below 
  the CB edge, suggesting it can accept one electron. However, as seen in 
  Fig.~\ref{fig:dos-alzn-oi}, the charged  state  $(\mathrm{Al_{Zn}-O_{i(o)}})^{-}$ is not spin-polarized and the in-gap states have energies between the VBM and the middle of the gap. 
  The combination where oxygens take up tetrahedral positions  is unstable and 
  relaxes to a variant of split configuration, where the O$_2$ dimer 
  is located over the Al in the adjacent ZnO layer. 
  This configuration, named $\mathrm{Al_{Zn}-O_{i(s)}}$ in 
 Fig. \ref{fig:dos-modelos}(8),  has 
  an excess of  0.14 eV in its formation energy compared to the combination where oxygen 
  is in an octahedral site. Its DOS in both neutral (Fig. \ref{fig:dos-modelos}(8)) and charged states are similar to the case of octahedral interstitial, and will  not be further discussed.
 
 \begin{figure}[htbp]
\begin{center}
\includegraphics[width=7.0cm]{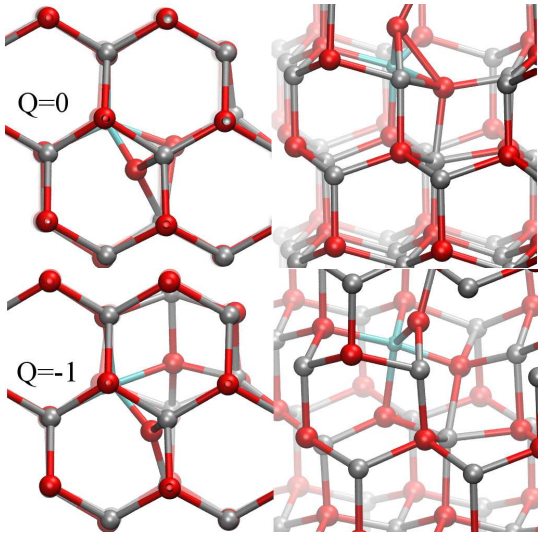}
\caption{(Color online) Views of defect $\mathrm{Ga_{Zn}-O_{i(o)}}$. Top: Neutral state. Bottom: charged state with one excess electron. Cyan, red, and gray balls represent Ga, O, and Zn atoms, respectively.}
\label{fig:defect-oi-gazn}
\end{center}
\end{figure}
 
 \begin{figure}[htbp]
\begin{center}
\includegraphics[width=7.0cm]{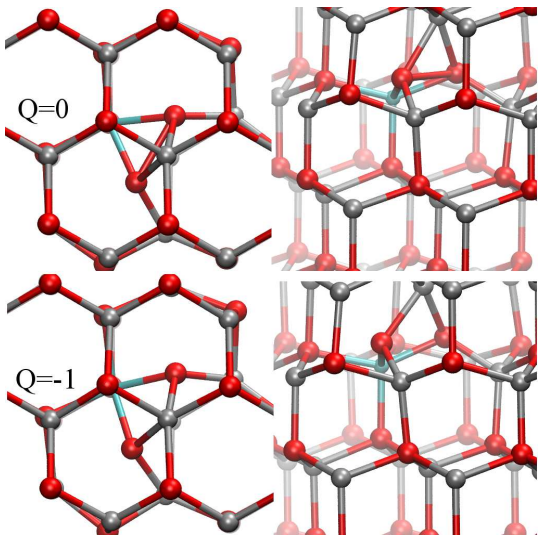}
\caption{(Color online) Views of metastable defect combination $\mathrm{Ga_{Zn}-O_{i(o)}*}$. Top: Neutral state. Bottom: state charged with one electron. Cyan, red, and gray balls represent Ga, O, and Zn atoms, respectively.}
\label{fig:defect-oi-gazn-meta}
\end{center}
\end{figure} 
     
  Results are very similar for the combination $\mathrm{Ga_{Zn}-O_{i(o)}}$ in the
   neutral and the singly charged states (Figs.~\ref{fig:dos-gazn-oi}(1, 2) and 
   Fig.~\ref{fig:defect-oi-gazn}). 
 The doubly charged state $\mathrm{(Ga_{Zn}-O_{i(o)})}^{2-}$ has been found to be thermodynamically stable\cite{demchenko2011} for values of the 
 Fermi level at 0.4 eV below the CBM and higher. However, according to 
  our electronic calculations the Fermi level would be  over the CBM, and the highest occupied level is the 
 lowest conduction band. This electronic structure is not consistent with a transition level below the CBM.  For Fermi level over the CBM, as in our  Ga-
 ZnO,  the state is plausible and deserves to be considered. Its DOS is shown in Fig. \ref{fig:dos-gazn-oi}(3), where it can be seen that the in-gap defect 
 levels are close to the VBM, a situation similar to the singly charged defect. Therefore, it seems that the doubly charged state cannot give rise to the HAXPES 
 peak near the CBM. 
 
 We have found a metastable configuration of $\mathrm{(Ga_{Zn}-O_{i}*)}$ that shows a DOS with a peak near the CBM, which meets the HAXPES
 data,  as it can be seen in Fig. \ref{fig:dos-gazn-oi}(4-6). 
 This configuration can be appreciated in Fig. \ref{fig:defect-oi-gazn-meta}. Its main difference with the stable configuration shown in 
 Fig.~\ref{fig:defect-oi-gazn} is a rotation of the  O$_2$ dimer.  
 The formation energy differences of this configuration with respect to the stable one are 0.25, 0.79, and 0.76 eV for the neutral, 
 single-charged and double-charged states, respectively.

\begin{table*}
\caption{ \label{tab:resumen}
Possible defects and charge states $0\le\alpha\le 1$. Positive (negative) $\alpha$ indicate depopulation (population) of donor (acceptor) bands. }
\begin{ruledtabular}
\begin{tabular}{lll}
Ga-ZnO & Al-ZnO  & Property \\
\hline
$\mathrm{Ga_{Zn}}$ &  & Stability, electron donation, band-filling. \\
 & $(\mathrm{Al_{Zn}})^{+}$ & Stability, electron donation. \\
$\mathrm{(Ga_{Zn}-V_{Zn}})^{-}$ & $\mathrm{(Al_{Zn}-V_{Zn}})^{-}$ & Compensation. \\
$\mathrm{(Ga_{Zn}-O_{i}}*)^{-}$ & $\mathrm{(Al_{Zn}-O_{i}})^{-}$  &  Compensation, HAXPES band. \\
 & $(\mathrm{Al_{i(o)}})^{+}$ & HAXPES band, electron donation.  \\
\end{tabular}
\end{ruledtabular}
\end{table*}
  
 It must be mentioned that our previous analysis of photoemission spectra by means of
  comparison with calculated DOS do not account for electronic relaxation 
 in the final state. This is particularly important in the cases of Fig.~\ref{fig:dos-alzn-oi} and Fig.~\ref{fig:dos-gazn-oi}(1-2). In these cases, 
 there is a strong variation in the energy of top last occupied orbital of the charged state, when it gets emptied by the photoemission. 
 This abrupt change deserves further analysis.  The HAXPES binding energies are in fact excitation energies or 
 quasiparticle energies. Due to the size of our system it is not possible to perform  a quasi-particle calculation. 
 However, the excitation energy can be estimated using 
 the Slater-Janak theorem as established in Ref.~\onlinecite{olovsson05}. The binding energy of the top-most valence  
 electron $E_B$ in the charged state ($N+1$ electrons) can be obtained as 
 \begin{equation}
 E_B= E_{N}-E_{N+1} \simeq  -\varepsilon_i(1/2) \simeq  -\frac{1}{2} [ \varepsilon_{i}(0) + \varepsilon_i(1) ].
 \end{equation}
where  $\varepsilon_i(\eta_i)$ is the energy of the orbital that becomes unoccupied in the excitation, and $\eta_i$ is the occupation number. 
We have obtained the values $E_B=-1.80$ and  -1.64 eV  for  $(\mathrm{Al_{Zn}-O_{i}})^{-}$ and $(\mathrm{Ga_{Zn}-O_{i}})^{-}$, respectively. 
Graphically, the HAXPES emission peak should be halfway between the energies 
of level that is filled at the charged state ($Q=-1$) 
and empty at the uncharged state ($Q=0$). 
Here the one-electron energies are referred to the supercell average electrostatic potential as 
usual in periodic DFT calculations. Adding the Fermi level energy one recovers the usual binding energy.  
In the same calculation the CBM is at 2.22 eV. Hence, if the material is n-type and the Fermi level is just below the CBM, as in Al-ZnO,
  the binding energy would be 0.42 eV below the Fermi level. Hence, the defect $(\mathrm{Al_{Zn}-O_{i(o)}})$ can still be responsible of the small HAXPES
  peak. In the case of Ga-ZnO, for the stable $(\mathrm{Ga_{Zn}-O_{i(o)}})$, the HAXPES peak would be shifted by 0.16 eV towards the valence band, and considering that the Fermi level is also shifted by 0.4 eV in the opposite direction, we think this separation would result in a double peak structure, in 
  disagreement with the experiment. However, the metastable configuration has a band that overlaps with the CB edge, in nice agreement with 
  the HAXPES spectrum. The issue of the relative stabilities of both configurations needs to be clarified in future work, using methods more 
  accurate for the total energy, obtaining the energy barrier between both configurations, and studying the effect of stress. 
  
Table \ref{tab:resumen} shows the defects and combinations that match the experimental data. 
  In Al-ZnO, the electrical properties are explained by substitutional $\mathrm{Al_{Zn}}$ compensated by Zn vacancies, in ratio 2:1  
  (we do not count  the Zn vacancy of  $\mathrm{Al_{Zn}}$), or combinations of both defects.  The small HAXPES band below the Fermi level can be attributed to a small amount of oxygen interstitials $\mathrm{Al_{Zn}-O_{i}}$ (exchanged by Zn vacancies) or interstitials $\mathrm{Al_{i}}$  (exchanged by $\mathrm{Al_{Zn}}$).  
  
 In Ga-ZnO, the electrical metallic behavior is explained by substitutional $\mathrm{Ga_{Zn}}$, partially compensated by acceptor defects. 
 $\mathrm{Ga_{Zn}-O_{i}}$ can provide compensation by accepting one electron, and it also would cause the observed HAXPES band below the 
 Fermi level. Compensation can also be achieved by Zn vacancies (or the combination $\mathrm{Ga_{Zn}-V_{Zn}}$), but it cannot be the cause of the 
 HAXPES band. An alternative to simple $\mathrm{Ga_{Zn}}$ is the combination $\mathrm{(Ga_{i(t)}-V_{Zn}-Zn_{i(o)}})^{+}$, which justify 
 simultaneously the HAXPES band and the electron donation. 
  
The presence of a HAXPES band below the Fermi level has been reported for Al-ZnO\cite{baoe2011}, Sn-In$_2$O$_3$\cite{korber2010}, and seems to be a robust effect in heavily doped semiconductor oxides. 
The sensitivity of HAXPES to CB states is facilitated by the fact that photoionization cross sections decrease much faster for O 2p shell than than for other elements, at photon energies over 2 keV. This factor reduces the signal from the VB edge, composed mostly of O 2p states, compared to the CB edge, that presents contributions from other atomic shells. The role of photoionization cross sections in the shape of the HAXPES spectra from the valence bands have been clearly shown.\cite{korber2010,panaccione2005} It is less clear for the CB edge, 
 as these states should be qualitatively different to the atomic levels. In fact, the HAXPES band was not observed for undoped, but still n-type, In$_2$O$_3$, similarly to our case. 
  For heavily doped Al-ZnO,\cite{baoe2011}  a coincident HAXPES spectra was obtained, although with a shorter VB tail than in the samples here studied. Moreover, transient capacitance 
 measurements revealed  the presence of a deep defect level with energy about 0.3 eV below the CBM, and concentration comparable to the shallow donor concentration.  These data could be explained by the presence  of 
 $(\mathrm{Al_{i(o)}})^{+}$, $\mathrm{(Al_{Zn}-O_{i}})^{-}$.

\section{Conclusions}
\label{sec:conclusions}

The differences between Al- and Ga- heavily doped ZnO  films  may have their origin in the association of each dopant with different structural defects in the ZnO matrix. In that way, a particular kind of defect could be present in the Ga-doped film, while absent, or in quite different concentration, in the Al-doped film. DFT and beyond calculations of the DOS induced by a list of defects and defect combinations around the band gap for these doped films have allowed to identify the most probable ones in each material. The comparison of the calculated DOS with the experimental HAXPES spectra and the analysis of the consequences that each particular defect or defect combination would have on the electrical and optical properties allowed to discard the non suitable ones. It seems that the Al cations can be located either at substitutional or octahedral interstitial  sites, while the Ga cations can be only  in substitutional sites. Al and Ga substitutional impurities $\mathrm{M_{Zn}}$  (M=Al, Ga) are donors but the electrical behavior of the Al-ZnO and Ga-ZnO films suggests that they cannot be the only doping induced defect in the ZnO matrix. Moreover,  different degrees of electron compensations are required in each material in order to explain the differences in carrier concentration and resistivity. The only acceptor combinations involving substitutional cations $\mathrm{M_{Zn}}$ are the complexes with Zn vacancies ($\mathrm{M_{Zn}-V_{Zn}}$ , M=Al, Ga ) and with oxygen interstitials $\mathrm{M_{Zn}-O_{i(o)}}$. For the Ga-doped film, the charged Ga complex with oxygen interstitial $(\mathrm{Ga_{Zn}-O_{i(o)}})^{-}$ may explain the HAXPES peak observed below the CB edge, although only as a metastable state with the same geometry as the uncharged state. The uncharged state presents a half occupied band that may explain the metallic behavior of Ga-ZnO. For the Al-ZnO film, interstitial $(\mathrm{Al_{i(o)})^{+}}$ is the best candidate to explain the small HAXPES peak observed near the Fermi level, but some other defects would be present in the Al-doped film. The acceptors $\mathrm{Al_{Zn}-V_{Zn}}$ and/or $\mathrm{(Al_{Zn}-O_{i(o)})^{-}}$ can explain the compensation and the resistivity semiconducting behavior, and the second can also contribute to the HAXPES band near the Fermi level. 

\acknowledgments

This work was supported by the European Project  NANOCIS  of the FP7-PEOPLE-2010-IRSES.
The authors thankfully acknowledge the computer resources, technical expertise and assistance provided by the 
Madrid Supercomputing and Visualization Center (CeSViMa) and the Barcelona Supercomputing Center (BSC). The authors acknowledge M. Gab\'as for many stimulating discussions and for allowing to use her HAXPES data prior to publication.  
 

\begin{thebibliography}{35}
\bibitem[{hax()}]{haxpesfit}
\bibinfo{note}{The HAXPES signal around the Fermi level from a gold film grown
  over the Ga-ZnO has been fitted by a complement error function
  $A\times\mathrm{erfc}[(E-E_F)/\sqrt{2}\sigma]$ with $\sigma=0.206$ eV.}
\bibitem[{ali()}]{alignment}
\bibinfo{note}{In the conventional practice of setting the energy zero at the
  Fermi levels, these 0.35 eV are the difference between the binding energies
  of the VBM of Ga-ZnO and Al-ZnO.}
\end{thebibliography}

\end{document}